\documentclass[sigconf, pbalance=True]{acmart}
\AtBeginDocument{%
  }

\setcopyright{none}
\renewcommand\footnotetextcopyrightpermission[1]{}
\RequirePackage[
  datamodel=acmdatamodel,
  style=acmnumeric,
  ]{biblatex}

\usepackage{gensymb}
\usepackage{enumitem}

\begin{document} 

% Title
\title{SARLink: Satellite Backscatter Connectivity using Synthetic Aperture Radar}

% Authors
\author{Geneva Ecola}
\orcid{0009-0006-6490-6778}
\affiliation{%
    \institution{Stanford University}
    \city{Stanford}
    \state{CA}
    \country{USA}}
\email{gecola@stanford.edu}
\author{Bill Yen}
\orcid{0009-0007-8539-9223}
\affiliation{%
    \institution{Stanford University}
    \city{Stanford}
    \state{CA}
    \country{USA}}
\email{billyen@stanford.edu}
\author{Ana Banzer Morgado}
\orcid{0000-0002-6988-9974}
\affiliation{%
    \institution{Stanford University}
    \city{Stanford}
    \state{CA}
    \country{USA}}
\email{morgado@stanford.edu}
\author{Bodhi Priyantha}
\orcid{0009-0002-5652-3161}
\affiliation{%
    \institution{Microsoft}
    \city{Redmond}
    \state{WA}
    \country{USA}}
\email{bodhip@microsoft.com}
\author{Ranveer Chandra}
\orcid{0000-0002-4175-1404}
\affiliation{%
    \institution{Microsoft}
    \city{Redmond}
    \state{WA}
    \country{USA}}
\email{ranveer@microsoft.com}
\author{Zerina Kapetanovic}
\orcid{https://orcid.org/0000-0001-6240-5511}
\affiliation{%
    \institution{Stanford University}
    \city{Stanford}
    \state{CA}
    \country{USA}}
\email{zerina@stanford.edu}

%Short author list
\renewcommand{\shortauthors}{Ecola et al.} 

%Keywords
\keywords{Satellite, Backscatter Communication, Synthetic Aperture Radar, Passive, Sentinel-1, Subaperture Processing, Remote Connectivity}

% Abstract
\begin{abstract}
SARLink is a passive satellite backscatter communication system that uses existing spaceborne synthetic aperture radar (SAR) imaging satellites to provide connectivity in remote regions. It achieves orders of magnitude more range than traditional backscatter systems, enabling communication between a passive ground node and a satellite in low earth orbit. The system is composed of a cooperative ground target, a SAR satellite, and a data processing algorithm. A mechanically modulating reflector was designed to apply amplitude modulation to ambient SAR backscatter signals by changing its radar cross section. These communication bits are extracted from the raw SAR data using an algorithm that leverages subaperture processing to detect multiple bits from a target in a single image dataset. A theoretical analysis of this communication system using on-off keying is presented, including the expected signal model, throughput, and bit error rate. The results suggest a 5.5~ft by 5.5~ft modulating corner reflector could send 60 bits every satellite pass, enough to support low bandwidth sensor data and messages. Using Sentinel-1A, a SAR satellite at an altitude of 693~km, we deployed static and modulating reflectors to evaluate the system. The results, successfully detecting the changing state of a modulating ground target, demonstrate our algorithm's effectiveness for extracting bits, paving the way for ultra-long-range, low-power satellite backscatter communication.
\end{abstract}

%Builds Document
\maketitle
\pagestyle{plain}
\section{Introduction}
\label{sec:Introduction}

% Introduce expansion of satellites and future challenges
There are over 9,100~active satellites in orbit today, and this number is projected to increase to tens of thousands in the coming years~\cite{Feder_2022_SatelliteProjection, ESA_2024_SpaceDebrisReport}. These satellite systems enable a wide variety of applications, such as Internet connectivity and imagery of remote regions of the Earth, but their proliferation has also created new challenges. This includes exacerbating issues of spectrum efficiency and co-existence with existing networks, obstruction of astronomical observations, increased light pollution, and generation of more space debris~\cite{Hall_2019_SatelliteThreats}. These challenges have motivated recent efforts to improve spectrum and hardware efficiency in space by developing new joint sensing and communication techniques. With this in mind, this paper investigates whether existing remote imaging satellites that use Synthetic Aperture Radar (SAR) can also support wireless communication without modifying their existing infrastructure.

\begin{figure}[t]
\centering
\includegraphics[width=0.5\textwidth]{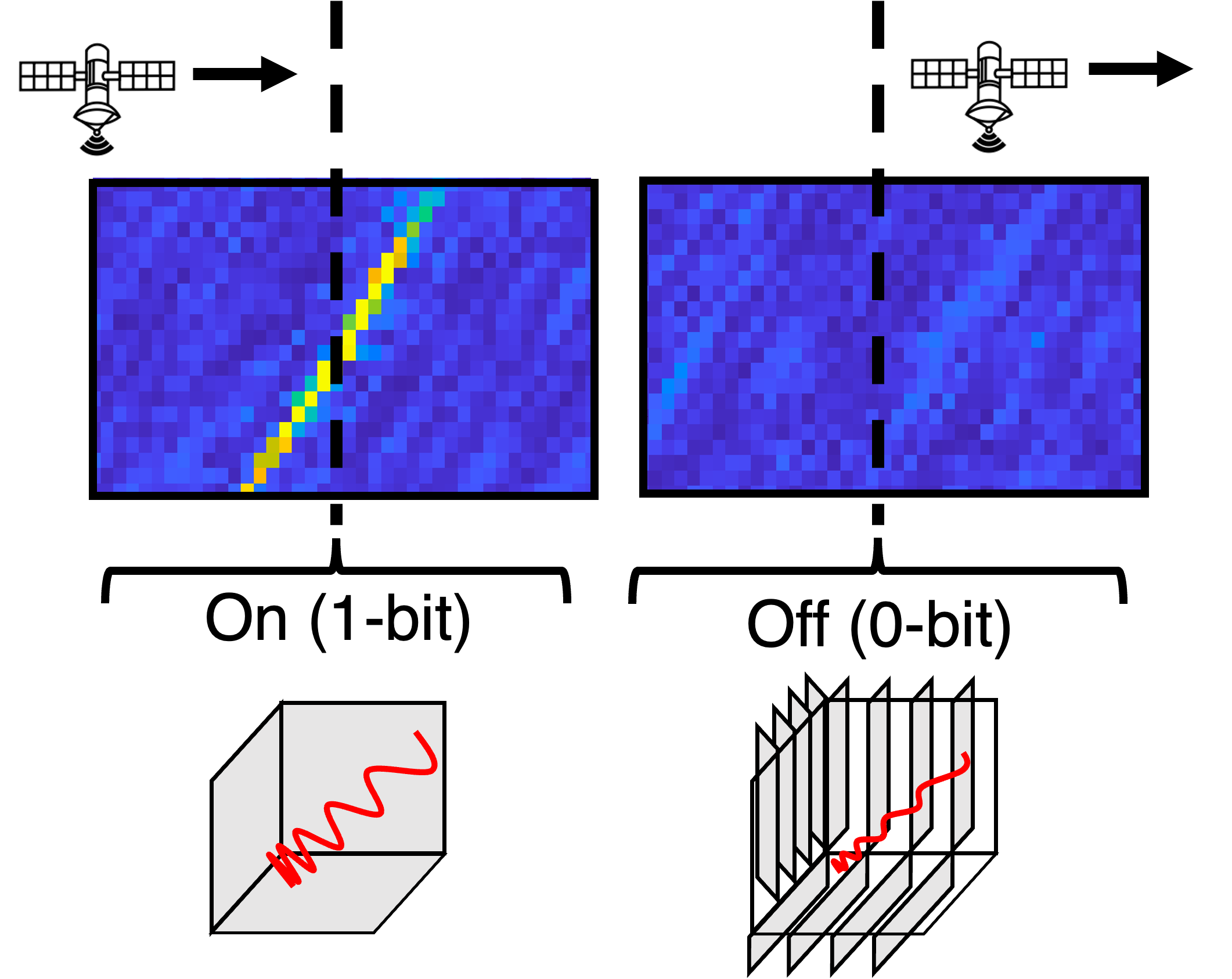}
\captionsetup{skip=1pt}
\caption{The corner reflector modulates incoming SAR signals by changing its radar cross section as the satellite images.}
\label{fig:High Level Idea}
\Description{Diagram of detecting one and zero bits using a mechanically modulating corner reflector. The reflector can send a one or a zero bit by changing its radar cross section at different points in time while the satellite images it.}
\end{figure}

% Previous work on SAR communications
SAR satellites, typically located in low earth orbit (LEO) at altitudes of 300 to 2000~km, are designed to generate images of the Earth's topography by transmitting radio frequency (RF) signals at the terrain and observing the received backscatter~\cite{LEO, Curlander_1992_SARSystems}. SAR has long been recognized as a promising technology for integrated sensing and communication, and there have been many efforts to incorporate communication into these systems. Most previous research requires modification of the SAR signal itself, which makes these techniques incompatible with current satellites~\cite{Wang_2019_JointSAR, Zheng_2024_JointSAR, Herschfelt_2017_JointSAR}. Prior work that does propose using existing systems necessitates transmitting active signals back to the satellite, which is often not possible due to spectrum allocation regulations and so far has only been demonstrated in simulation~\cite{Piccioni_2021_SDRGroundTarget, Piccioni_2025_JointFramework}. Using an existing SAR satellite for communication is a major challenge because of the inability to control the transmitted signal, change the receiver hardware, or actively transmit signals to the satellite, all while trying to establish a communication link with devices in LEO. In this paper, we overcome these challenges and present SARLink, a system that enables passive ground-to-satellite connectivity using existing satellite infrastructure.

% Our approach to SAR backscatter
SAR systems receive weak backscatter signals from passive terrain, and SARLink leverages this capability to detect passive modulation of these ambient signals. Our approach is analogous to short-range RFID systems, where the SAR satellite acts as the reader for a passive on-ground device that modulates a carrier signal \cite{Want_2006_RFID}. Despite the similarities, using SAR systems for wireless communication is not straightforward. SAR processing is complex since the radar captures data as the satellite moves thousands of meters per second in orbit. Moreover, existing processing techniques are specifically designed for imaging, where it is assumed that the measured terrain remains static during data collection~\cite{Curlander_1992_SARSystems}. Therefore, extracting information bits from a cooperative target requires a new processing technique to detect modulation from a single image.

% SARLink approach
Our key insight is that SAR images can be generated from subsets of the full image data to capture a target's state at different points in time. We theoretically analyze this technique and show that information bits can be embedded and extracted from SAR image datasets using on-off keying (OOK) by modulating a ground target's radar cross section (RCS). We took inspiration from SAR calibration techniques that use corner reflectors because of their large RCS and designed a mechanically modulating corner reflector to evaluate the feasibility of SARLink~\cite{australiaCR}. \autoref{fig:High Level Idea} illustrates how the reflector switches between highly reflective and non-reflective states to send information. This proof-of-concept device, while not low power, allowed us to deploy the system and show that the proposed processing was an effective method for backscatter communication.

\begin{figure}[t!]
\centering
\captionsetup{skip=1pt}
\includegraphics[width=0.5\textwidth]{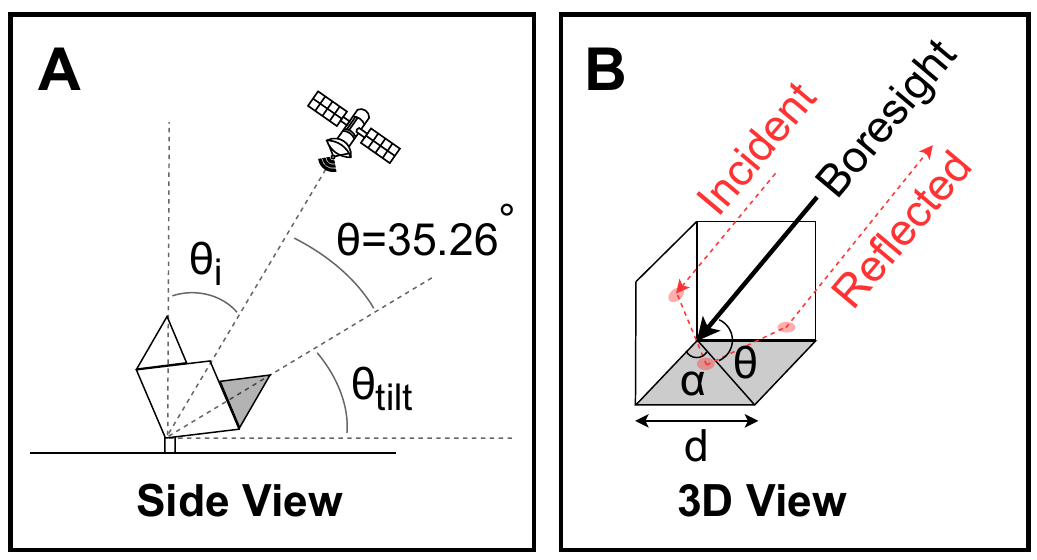}
\caption{The side view (A) shows the corner reflector's boresight aligned with the incident SAR signals by tilting it at angle $\boldsymbol{\theta_{tilt}}$ to maximize its RCS based on the incidence angle, $\boldsymbol{\theta_{i}}$. The 3D view (B) shows how when $\boldsymbol{\alpha=45\degree}$ and $\boldsymbol{\theta=35.26\degree}$ the boresight of the reflector is aligned with incident waves and redirects them back to their source with maximum power.}
\label{fig:background_CR}
\Description{Two images, a side view and a 3D view, of corner reflectors depict their three perpendicular panels that form a cube to reflect the signal.}
\end{figure}

\begin{figure*}[t!]
\centering
\captionsetup{skip=1pt}
\includegraphics[width=1\textwidth]{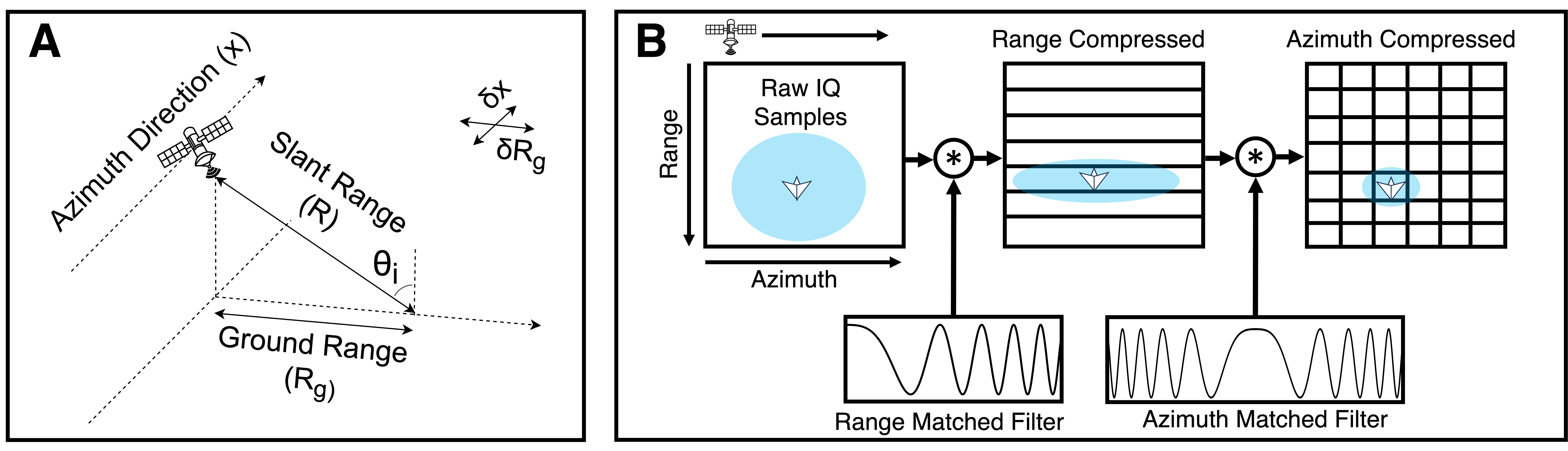}
\caption{(A) The satellite moves in the along-track or azimuth direction and images targets on the ground, whose range can be defined using either the slant or ground range geometry and are related using the incidence angle, $\boldsymbol{\theta_{i}}$. (B) SAR processing takes the raw IQ samples and convolves them in the range direction with a matched filter and then again in the azimuth direction with another matched filter. This process compresses the data into range and azimuth bins that form the final pixels of the resolved SAR image.}
\label{fig:background_SAR}
\Description{The figure is split into two subfigures, A and B. Subfigure A shows the geometry of SAR imaging where the satellite looks to its side at look angle theta_look and incidence angle theta_i to image a swath. Subfigure B summarizes SAR processing and shows how a radar target compresses as the processing in the range and azimuth are completed and the pixels of the final image are created.}
\end{figure*}

% Applications
SARLink has the potential to expand the deployment range of wireless sensors by backhauling data from remote regions that lack ground-based communication infrastructure, such as mountains, deserts, and oceans. This sensor data could aid remote sensing models that rely on ground-truth sensor data to train and verify their performance, such as those using SAR data itself~\cite{Wig_2024_inSARmodel}. As the first technique for passively sending information bits from the ground to a SAR satellite — and with some SAR systems offering open-access data — this system could enable anyone to send information without expensive licenses or subscriptions. Thus, it provides an accessible way of sending messages in areas without connectivity or in censored environments where active radio transmissions cannot be used. Furthermore, SARLink requires no modification of the satellite infrastructure, either on the transmit or receive side, so it is compatible with existing SAR systems, such as RADARSAT, ICEYE, or Umbra~\cite{iceye, umbra, radarsat}. We demonstrate our system using the European Space Agency (ESA) satellite Sentinel-1A, as the data is freely available and the system regularly images all the land on Earth. In addition, SARLink can be used with upcoming SAR satellites, including Sentinel-1C, Sentinel-1D, and NASA's NISAR mission~\cite{Sansosti_2104_2ndGenSAR, nisar, sentinellaunch}.

Our key contributions are summarized below:
% List of these key contributions
\begin{itemize}[leftmargin=*, topsep=3pt] % Adjust the space before the list
\item [(1)] We present a method for passive satellite backscatter connectivity using SAR satellite technology that enables a device on the ground to send information to a satellite without actively transmitting RF signals.
\item [(2)] We design a new SAR processing technique, leveraging subaperture processing, for wireless communication that extracts information bits from changes in a target's reflectivity during a single satellite pass using OOK. We implement this algorithm using a real SAR processor for Sentinel-1 SAR data. 
\item[(3)] We propose a theoretical model that estimates 60 bits can be sent every satellite pass with a bit error rate (BER) of 1\% using a modulating 5.5~ft by 5.5~ft square trihedral corner reflector.
\item [(4)] We designed a mechanically modulating corner reflector as a proof of concept ground target that passively applies amplitude modulation to SAR signals. We used this device to conduct field experiments with Sentinel-1A. We demonstrated that the processed results match the theoretical model and that it can detect state changes in the reflector during a satellite pass.
\end{itemize}
\section{Background}
\label{sec:Background}

% Introduction to background topics
This section introduces the key technologies used to enable SARLink. We describe the properties of corner reflectors that make them an effective method for creating bright targets in SAR data. We also provide an overview of SAR processing and discuss the properties of SAR images that impact how our communication system works. 

\subsection{Corner Reflector Theory}\label{background:CR}
Corner reflectors, shown in ~\autoref{fig:background_CR}, are retroreflectors that passively redirect RF waves back to their emission source and have extremely large RCSs for their physical size. The square trihedral corner reflector consisting of three square planes has the largest maximum RCS, $\sigma_{max}$, among typical reflector designs,
\begin{equation} \label{eq:CRrcsmax}
\sigma_{max} = \frac{12\pi d^{4}}{\lambda^2},
\end{equation}
where $d$ is the side length of the square panel of the reflector, and $\lambda$ is the wavelength of the incoming signal. The maximum RCS occurs when the incidence angle is parallel to boresight, meaning that $\alpha$=45\degree~and $\theta$=35.26\degree ~\cite{Doerry_2008_SARReflectors}. Therefore, depending on a satellite's orbit, the reflector will need to be angled correctly to achieve this maximum RCS.

\subsection{Synthetic Aperture Radar}
\label{subsec:SARBackground}
SAR transmits RF chirps to the ground at a specified pulse repetition frequency (PRF) and measures the backscatter as the radar moves rapidly across the imaging area. The backscattered signals will have varying amplitudes and phases depending on the nature of the terrain. Processing must be done to resolve these complex received samples and convert them into an image. Resolution is the primary performance metric of a SAR system as it defines the minimum required distance between two points to distinguish them, which informs the number of pixels used in the final image. SAR uses two processing techniques to resolve targets in the range (across-track) and azimuth (along-track) directions. The geometry of SAR imaging is summarized in \autoref{fig:background_SAR}A.

Range processing uses samples received from a single transmitted radar pulse. The radar resolves samples based on their time of flight, and pulse compression is employed to improve the signal-to-noise ratio (SNR), as depicted in the first step of \autoref{fig:background_SAR}B. The range-compressed image will appear smeared in the along-track direction as it has not yet been azimuth compressed. Using this processing, the ground range resolution, $\delta R_{g}$, will be given by
\begin{equation} \label{eq:rangeresSAR}
\delta R_{g} = \frac{c}{2B\sin(\theta_i)},
\end{equation}
where $c$ is the speed of light, $B$ is the bandwidth of the radar chirp, and $\theta_i$ is the incidence angle (the angle between the radar beam and the normal to the earth's surface) as shown in ~\autoref{fig:background_SAR}A \cite{Curlander_1992_SARSystems}. The range resolutions of notable satellite SAR systems vary from 1 to 75~m depending on the imaging mode ~\cite{Sansosti_2104_2ndGenSAR}.

\begin{figure*}[th!]
\centering
\captionsetup{skip=1pt}
\includegraphics[width=1\textwidth]{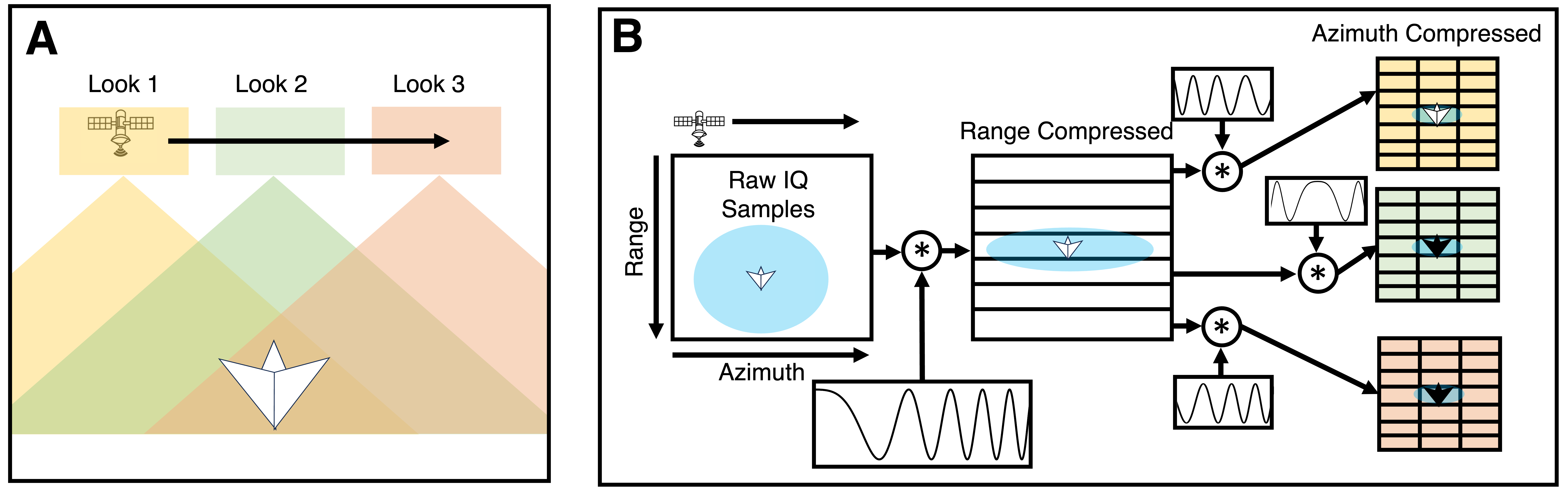}
\caption{Subaperture processing is used to enable backscatter communication in (A) each color represents a subset of data collected at different times during the satellite pass and is used to generate separate sublooks. (B) Subaperture processing uses the same range of compression processing but adapts the azimuth compression step to produce the desired number of sublooks.}
\label{fig:SARSublooks}
\Description{Diagram has two subfigures, A and B. A represents the different datasets collected during the satellite's trajectory as different colors, yellow, green, and red, to show how different points in the satellite's orbit are used to generate different looks. Figure B shows that subaperture processing can generate three images, again represented by yellow, green, and red, from these subsets of data using an adjusted matched azimuth filter for each.}
\end{figure*}

SAR gets its name from the way it handles azimuth processing. SAR processors use the radar platform's known movement to generate a synthetically large antenna and detect targets with better resolution than simpler side-looking airborne radar systems. The Doppler shift generated as the radar moves allows the system to resolve targets in the azimuth direction. Using this known frequency shift, pulse compression techniques, as shown in the second step of \autoref{fig:background_SAR}B, can be used to resolve the location of the point with an azimuth resolution, $\delta x$, that is only dependent on $L_{a}$, the antenna length \cite{Curlander_1992_SARSystems}
\begin{equation} \label{eq:azresSAR}
\delta x = L_a/2.
\end{equation}
The azimuth resolution is typically between 1 and 100~m for satellite SAR systems depending on the imaging mode ~\cite{Sansosti_2104_2ndGenSAR}. \autoref{fig:background_SAR}B is a simplified description of SAR processing; real data requires many other steps. Sentinel-1's detailed algorithm definition is an example of how complicated SAR processing can be~\cite{Piantanida_2019_DetailedAlg}.

In addition to enhanced azimuth resolution, SAR processors also have improved SNR due to their coherent processing of samples. The SNR of a single radar pulse is given by
\begin{equation} \label{eq:SLARSNR}
SNR_{o}=\frac{P_{t}G^2 A_{scat} \sigma^{o} \lambda^2}{(4\pi)^3 R^4 kT_{sys}B},
\end{equation}
where $P_{t}$ is the radar's transmit power, $G$ is the gain of the antenna, $A_{scat}$ is the scattering area, $\sigma^{o}$ is the normalized RCS, $R$ is the slant range from the target to the radar, $k$ is Boltzmann's constant, and $T_{sys}$ is the system noise temperature~\cite{Curlander_1992_SARSystems}. The SNR of a SAR system has additional gain from the range and azimuth processing. This gain is proportional to the number of coherently processed samples used to measure the target and is the product of the samples per radar pulse and the total number of pulses used to image it. The number of samples per pulse, $N_{R}$, will depend on the receiver bandwidth and $\tau_{p}$, the radar pulse duration\cite{Curlander_1992_SARSystems},
\begin{equation} \label{eq:RangeSampleNum}
N_{R} = B\tau_{p}.
\end{equation}
The total number of pulses, $N_{A}$, used to image a target will be a product of the time that it is in view of the radar's antenna beam, also called the radar's dwell time, and PRF of the radar, $f_{p}$,
\begin{equation} \label{eq:AzimuthSampleNum}
N_{A} = \frac{f_{p}R\lambda}{L_aV_{st}},
\end{equation}
where $V_{st}$ is the velocity of the satellite. The enhanced SNR of the final SAR image, $SNR^I$, will be
\begin{equation} \label{eq:SARImageSNR}
SNR^I = N_{R}N_{A}SNR_{o}.
\end{equation}
SAR systems can achieve significantly better azimuth resolution and SNR than noncoherent-based processing techniques, enabling them to take impressively detailed images of the Earth. 

When using SAR to view a specific object, it is also useful to define a metric to compare the signal strength of the target to that of other background scatterers. This is defined as the Signal-to-Clutter Ratio (SCR), which is the ratio of the desired signal from a target to signals from background objects. This value will depend on the target's radar cross section, $\sigma^{o}_{target}$, and the clutter, $\sigma^{o}_{clutter}$, within an image's resolution cell~\cite{Freeman_1992_SCRDef}
\begin{equation} \label{eq:SCR}
SCR=\frac{\sigma^{o}_{target}}{\sigma^{o}_{clutter}},
\end{equation}
and is useful to determine how easily a target will be distinguished from the background of an image.
\section{SAR-Based Backscatter Communication} 
\label{sec:theoretical}

Using a SAR system for communication requires embedding information bits into the image dataset while the radar passes overhead. As previously discussed, to make our system compatible with existing SAR satellites, we cannot change the radar's transmitter or receiver and also cannot transmit active signals to communicate. Therefore, our approach must modulate SAR signals passively in a way that can be extracted from the image data. In this section, we propose and derive the expected performance of a SAR backscatter communication system.

First, let's consider the simplest realization of this system. A cooperative target with two configurable RCS states is deployed, allowing the radar to detect whether the target is in its on-state (large RCS) or off-state (small RCS) each time it passes over the reflector. This approach has clear, practical limitations as only one bit of information would be sent every time the area is imaged. Instead, configuring the on-ground modulating device to change its RCS state multiple times during a pass would be more effective. However, sensing multiple state changes is non-trivial.

To detect these changes or, in other words, transmit multiple data bits per satellite pass, we must process the image data so that the RCS of the modulating device can be estimated at many different points in time during the satellite's imaging. A promising approach is subaperture processing, which is typically used to reduce speckle noise in SAR images~\cite{porcello_speckle_1976}. 
This process divides the "synthetic aperture" into smaller pieces to generate multiple lower-resolution images called sublooks and averages them into a final image. For our application, we can instead use each sublook separately to evaluate the state of our RCS modulating device many times during the satellite pass. This section will discuss the impact of subaperture processing and the expected performance as a technique for OOK backscatter communication.

\subsection{Subaperture Processing for Communication}
\label{subsec:subapertureforcomm}

Subaperture processing uses subsets of data from different points in the satellite trajectory as illustrated in ~\autoref{fig:SARSublooks}A. ~\autoref{fig:SARSublooks}B shows a basic overview of how the original SAR processing algorithm, from \autoref{fig:background_SAR}B, is adapted to achieve subaperture processing. After range compression, the processor uses a subset of the data and a modified matched filter to perform azimuth compression~\cite{Moreira_1992_SubapertureProcessing}. The sublooks capture the scene at different center times separated by $\Delta t$~\cite{ouchi_multilook},
\begin{equation} \label{eq:TimeDiff}
\Delta t=\frac{N_{A}}{mf_{p}},
\end{equation}
where $m$ is the number of equally sized sublooks generated. This time difference will define the sampling frequency of the communication system and thus determine the maximum modulation speed of the on-ground device.

Subaperture processing has two major impacts on the output images: the image resolution and the SNR. Since we are working with fully range compressed data, the range resolution of the image remains the same. However, we are reducing the length of the synthetic aperture, which degrades the azimuth resolution from \autoref{eq:azresSAR} to
\begin{equation} \label{eq:azres_sublook}
\delta_{x,sublook}=\frac{mL_{a}}{2}.
\end{equation}
Since the resolution cell of the image increases with the number of sublooks, there will also be background clutter from other locations averaged along with the modulating device. 

The other trade-off in increasing the number of sublooks is a reduction in the signal power and thus the SNR of the image. Using ~\autoref{eq:SARImageSNR}, we find that since subaperture processing effectively reduces the number of coherent samples used to process the image, it also reduces the SNR,
\begin{equation} \label{eq:SublookSNR}
SNR^{I}_{L} = \frac{SNR^{I}}{m}.
\end{equation}

With this in mind, we can determine the expected signal model of our communication system. First, let's consider the power received from a SAR image resolution cell without using subaperture processing. The received power, y, will depend on the target's RCS, the clutter's RCS, and $w$, the random noise from the receiver
\begin{equation} \label{eq:signalmodel_single}
y = \zeta(\sigma^{o}_{target} + \sigma^{o}_{clutter}) + w.
\end{equation}
The received power will be proportional to the RCS of both the target and the clutter multiplied by a constant $\zeta$, which is derived from the signal portion of \autoref{eq:SARImageSNR}.

This model will change depending on the number of sublooks due to the reduction in detected signal power. We represent the signal model of the communication system as the power measured at the resolution cell containing the modulating target, which is evaluated at $n$, the center times of each sublook
\begin{equation} \label{eq:signalmodel}
y[n] = \frac{\zeta(\sigma^{o}_{target}[n] + \sigma^{o}_{clutter}[n])}{m} + w[n].
\end{equation}
The model shows how the power is reduced for both the target and clutter signal compared to the receiver noise. To evaluate the system further, we need to make some assumptions regarding this model. First, we assume that the RCS of the background clutter remains constant across sublooks so that we can analyze the system simply as an OOK scheme which depends only on the RCS of the target. This assumption is reasonable in areas with very consistent clutter, like the ocean, but less so for areas with high clutter. Second, we assume that the corner reflector's RCS remains constant across sublooks in both its on and off states. In other words, when the target is in its on-state, the received power remains unchanged regardless of the sublook used. This assumption is valid in all spaceborne deployment scenarios, as the satellite observes the target from such a narrow angular range that the RCS of the target in its on and off states will appear consistent across sublooks~\cite{jauvin2019integration}.

\subsection{Bit Rate and Throughput}
\label{subsec:throughput}
The radar's PRF will set the system's maximum achievable bit rate. Since the satellite and ground target will not be synchronized, we need to ensure that at least two radar pulses are used to detect each state of the reflector. Thus, the maximum achievable bit rate will be $f_{p}/2$. Sentinel-1, for example, uses PRFs between 1 and 3 kHz, meaning that the theoretical maximum achievable bit rate will be between 500 to 1500 bps ~\cite{sentinelperformance}. As such, if the maximum number of sublooks is used where $m=N_A$, then the maximum achievable throughput for each satellite pass, $T$, will be $T=N_{A}/2$.~\autoref{eq:AzimuthSampleNum} shows that the throughput will thus be related to the satellite's velocity, its range from the target, and the ratio of the radar wavelength to its antenna length.

\subsection{Bit Error Rate} 
\label{subsec:biterrorrate}
SARLink uses OOK to communicate using SAR signals. The basic expression for the BER using OOK is given by~\cite{safak_digital_2017}
\begin{equation} \label{eq:BERBasic}
BER = Q\left(\sqrt{\frac{E_{b}}{N_{o}}}\right),
\end{equation}
where $E_{b}$ is the signal energy per bit, $N_{o}$ is the noise spectral density, and $Q(\cdot)$ is the tail probability of a standard normal distribution. $E_{b}/N_{o}$ will be half the difference between the reflector's SNR in its on and off states of each sublook, thus
\begin{equation} \label{eq:BERSAR}
BER = Q\left(\sqrt{\frac{SNR^{I}_{L,on}-SNR^{I}_{L,off}}{2}}\right),
\end{equation}
where $SNR^{I}_{on}$ is the SNR of the reflector in its on-state and $SNR^{I}_{off}$ is the SNR in its off-state. This analysis will be valid as long as in every sublook the SNR of the reflector in its on-state is greater than the radiometric accuracy of the radar. Otherwise, the state difference would not be detectable. To explicitly show how the number of sublooks impacts the BER, the expression can be rewritten in terms of the original SNR of the SAR image using \autoref{eq:SublookSNR}
\begin{equation} \label{eq:BERSAR_sigma}
BER = Q\left(\sqrt{\frac{SNR^{I}_{on} - SNR^{I}_{off}}{2m}}\right).
\end{equation}
From the signal model in \autoref{eq:signalmodel} we note that this expression also implies that if the clutter is constant in each sublook then it acts as an offset that cancels out and thus does not impact the BER of the system. This expression also gives intuitive insight into how to improve the performance of the system. If we assume that as the reflector size increases, its RCS in its on-state increases while its off-state remains relatively constant, then we find that a larger reflector will improve the system's BER.

\begin{figure*}[t!]
\centering
\includegraphics[width=\textwidth]{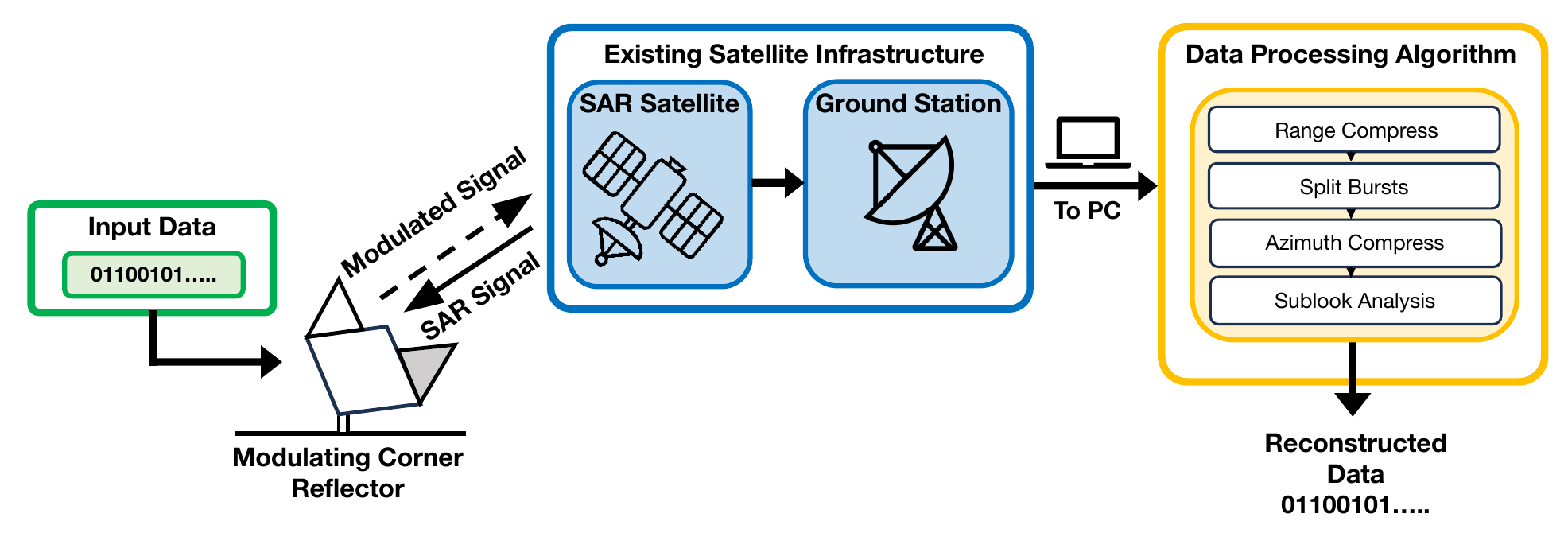}
\captionsetup{skip=1pt}
\caption{SARLink enables long-range backscatter communication using existing spaceborne SAR infrastructure by introducing a cooperative on-ground target as a modulating corner reflector. A standard laptop PC with access to the Internet can run the proposed data processing algorithm to extract the input data from the satellite's images.}
\label{fig:system-diagram}
\Description{A system overview shows a complete system diagram. The modulating corner reflector takes in arbitrary data and modulates the SAR signal. This is received by the SAR satellite infrastructure and sent to a ground station. From there, the data is accessible over the internet, a PC can be used to analyze the data.}
\end{figure*}
\section{SARLink System Design}
\label{sec:SARLink System Design}

SARLink is comprised of a data processing algorithm, SAR satellite, and modulating corner reflector (see~\autoref{fig:system-diagram}). 

\subsection{Data Processing Algorithm}
\label{subsec:ProcessorDesign}
\begin{figure}[t!]
\includegraphics[width=0.45\textwidth]{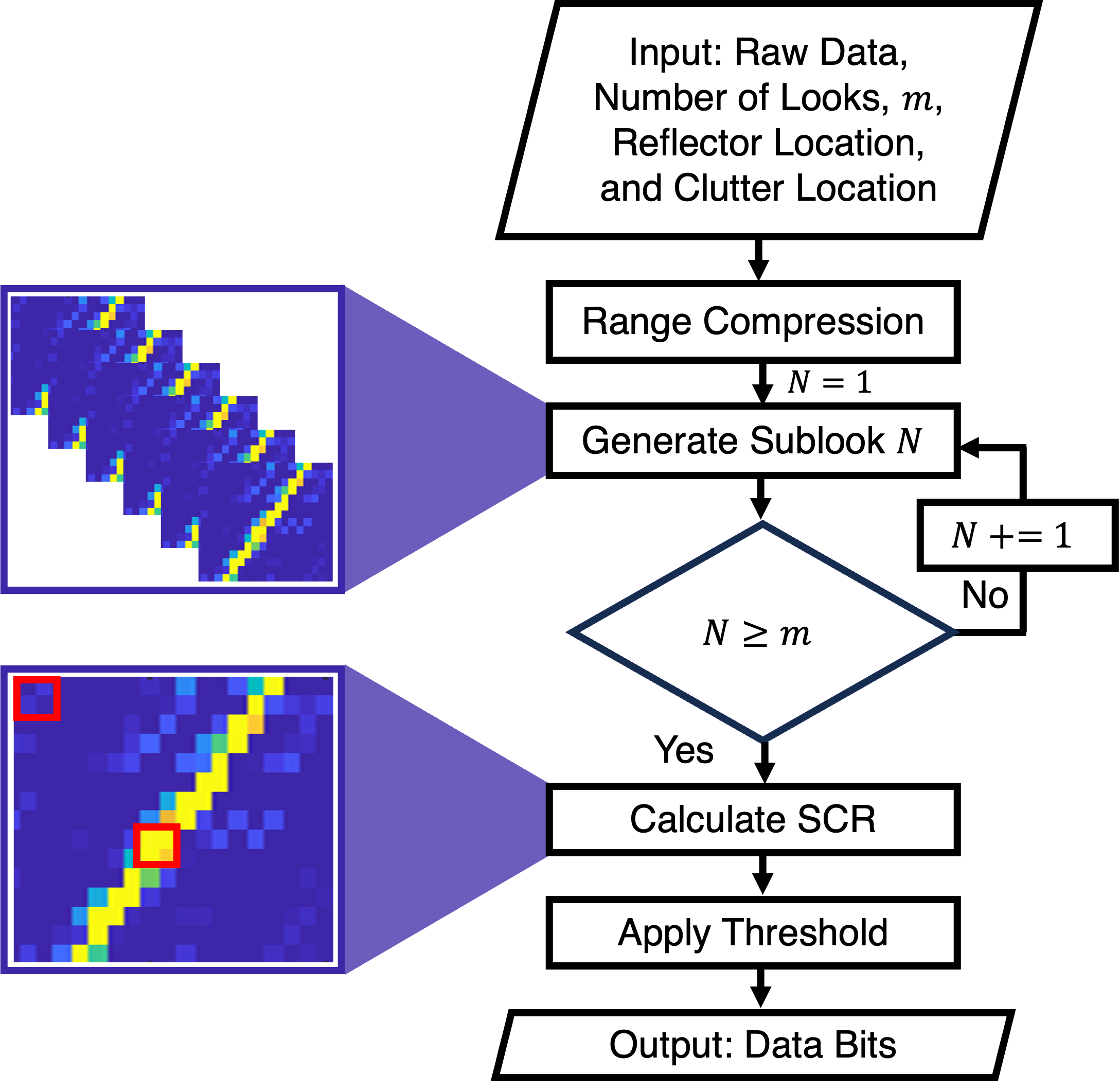}
\captionsetup{skip=1pt}
\caption{Diagram of the data processing algorithm shows how a SAR image dataset is processed into sublooks to obtain data bits.}
\label{fig:processing_diagram}
\Description{Diagram describes each of the algorithm's processing. The first input block has image data, number of looks, m, reflector location, and clutter location. Then, it points to the next step, which is range compression. Then there is a loop for generating sublooks until all m have been created. The next block says to calculate SCR, and an image is shown with red boxes around the target and the background clutter. The next step is to apply a threshold, and the final step is an output block that will provide data bits.}
\end{figure}

Existing SAR data processing techniques are not designed for communication. Therefore, we develop a new method of using subaperture processing to extract data bits. This data processing algorithm has four steps: range compression, burst splitting, azimuth compression, and image analysis of each sublook image to determine the state of the reflector as discussed in \autoref{sec:theoretical}. \autoref{fig:processing_diagram} is a data processing diagram that provides an overview of the algorithm. It also includes example images of SAR data of a corner reflector in sublook images; the reduction in azimuth resolution is evident by the smeared appearance of the reflector in the images.

 The inputs to the processing algorithm include the raw radar data, the number of sublooks required to process the data, the location of the reflector signal, and a nearby location of low clutter in the image. An existing SAR algorithm used to process images was modified and required modification to enable subaperture processing and subsequent detection of communication bits ~\cite{zebker_code}. The processor first fully range compresses the raw image data. Then the sublooks are generated. These sublooks are processed using $m$-independent equally sized subsets of the range compressed data. Once all of the sublooks are generated, the algorithm calculates the SCR using an equal number of pixels from the location of the reflector and a low clutter area as marked by the red squares. This is required to account for differences in amplitude of the measured signals due to differences in the antenna's gain. The signal is then reconstructed based on the detected RCS values and thresholded to generate the output data bits.

\subsection{SAR Satellite}
\label{subsec:SARSatellite}
SARLink does not require deploying new satellite infrastructure. Instead, it uses existing SAR satellites by modulating their backscattered signals and extracting the information sent by the reflector. This approach is fully passive and does not require any changes to the transmitted signal or receiver hardware. Therefore, SARLink is compatible with aircraft-based systems and existing satellite systems currently in orbit. The performance of the communication system will depend on specific radar parameters such as the PRF, and the theoretical analysis developed in this paper can be used to evaluate individual systems. The primary relevant difference between systems in terms of Hardware is the operating frequency, which may require changing the reflector size to acheive desired SNR values.

\subsection{Modulating Corner Reflector}
\label{subsec:ModulatingReflectorDesign}
The modulating corner reflector is a device that implements OOK by mechanically switching between a large RCS state and a small RCS state to send data bits. This is achieved by exploiting the geometry of a corner reflector. A corner reflector generates a large RCS using three orthogonal panels to bounce signals back to their source. Therefore, changing the angle of these panels relative to one another significantly affects its RCS. The contrast between perfectly aligned and misaligned panels is quite large; deviating the panel angle by just 2\degree~ can result in a transmission loss of 30 dB for large reflectors~\cite{anderson1985effect}. SARLink uses this sensitivity to create a mechanically modulating corner reflector that generates the desired high and low RCS states necessary for OOK.

\begin{figure}[t!]
\includegraphics[width=0.5\textwidth]{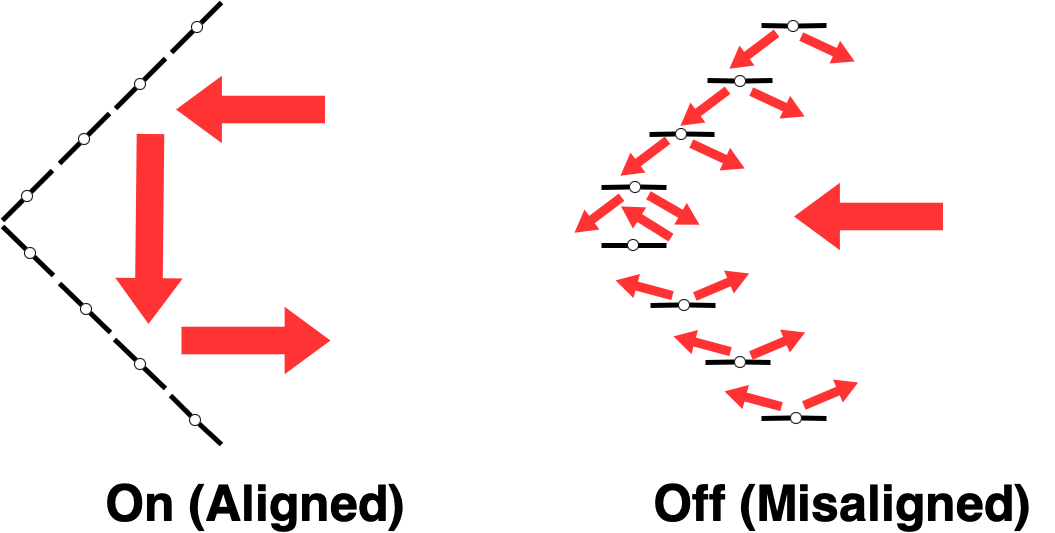}
\captionsetup{skip=1pt}
\caption{The mechanically modulating reflector scatters the incoming signals when the sub-panels are misaligned, resulting in a smaller RCS.}
\label{fig:mech_mod_overview}
\Description{The incoming radar signal scatters when the sub-panels are misaligned, resulting in less signal returning to the satellite and, therefore, a smaller RCS.}
\end{figure}

Our design splits each reflector panel into four sub-panels that can be tilted. When the sub-panels lay flush, the reflector behaves as a conventional corner reflector. When the reflector rotates its panels, they become misaligned such that the incoming signal is scattered in undesirable directions instead of back to the satellite, thus decreasing the RCS as illustrated in \autoref{fig:mech_mod_overview}. The modulating reflector has three main design parameters that will impact the overall performance of the communication system:

\textbf{Radar Cross Section.} The length of a corner reflector, $d$, as shown in \autoref{fig:background_CR}, should be at least 10 times the wavelength of the transmitted signal to be detectable by the radar~\cite{osman2014analysis}. For this system specifically, another consideration is the BER. \autoref{eq:BERSAR_sigma} shows that it depends on the RCS of the reflector in its on-state. The main trade-off with increasing the RCS is that scaling $d$ increases the device's size significantly.

\textbf{RCS States.} The difference in the reflector's RCS states determines the system's BER using OOK. Therefore, the design must maximize the difference between these two states as much as possible.

\textbf{Modulation Speed.} The chosen data rate will determine the required modulation speed of the reflector, which will fall within these general upper and lower bounds. The maximum required modulation speed, as discussed in ~\autoref{subsec:throughput}, will be the $f_{p}/2$, and the minimum will depend on the radar's dwell time, the amount of time that the radar spends imaging a specific location. The reflector must be able to modulate its state at least two times faster than the radar's dwell time to send more than one bit.
\section{Implementation}
To evaluate the effectiveness of subaperture-based backscatter communication, we designed a mechanically modulating corner reflector and demonstrated the system in the field using the ESA's Sentinel-1A satellite.

\subsection{Corner Reflectors}
We designed and implemented static and mechanically modulating corner reflectors to deploy during Sentinel-1A imaging.

\subsubsection{Static Reflectors}
\begin{figure}[t!]
\centering
\captionsetup{skip=1pt}
\includegraphics[width=0.5\textwidth]{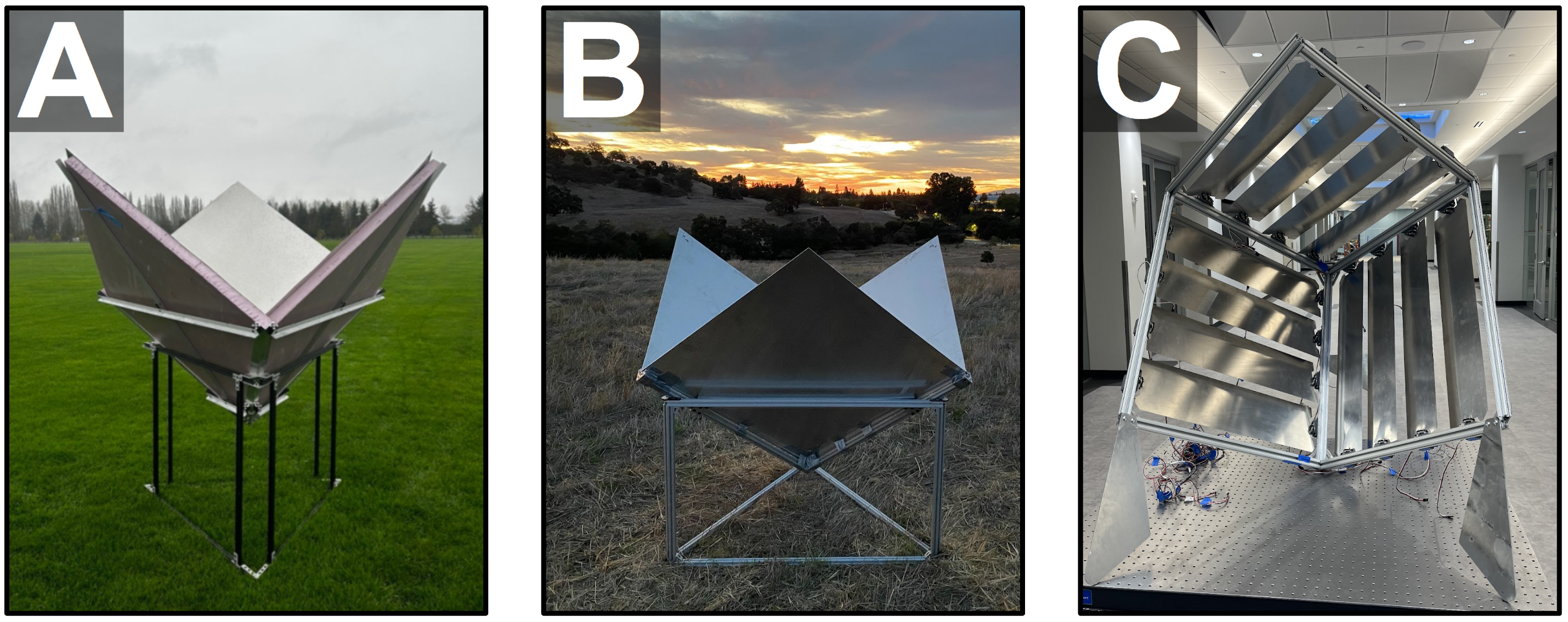} 
\caption{Reflector designs are shown including the (A) static 4~ft by 4~ft, (B) static 3~ft by 3~ft, and (C) mechanically modulating 2~ft by 2~ft.}
\label{fig:cornerreflectors}
\Description{Reflector designs are shown made of metal sheets and 80/20 extrusions. The modulating reflector is shown with its panels slightly tilted.}
\end{figure}

 We built three square trihedral corner reflectors with panel sizes of 2~ft by 2~ft, 3~ft by 3~ft and 4~ft by 4~ft and theoretical maximum RCSs of 32.4, 39.3 and 44.4~dB, respectively, at 5.4~GHz (see~\autoref{fig:cornerreflectors} A and B). The reflectors feature three square aluminum panels of their respective sizes and a frame that secures them to ensure their perpendicularity and elevate the panels from the ground. Among the different designs implemented, we found that 80/20 aluminum extrusions provided the best combination of strength-to-weight ratio and ease of assembly for the reflector frames~\cite{80_20_bars}. The frames are designed to be taken apart and assembled in outdoor deployments. 

\subsubsection{Mechanically Modulating Reflector}
To implement the mechanical modulation described in \autoref{subsec:ModulatingReflectorDesign}, a 2~ft by 2~ft reflector was constructed where each side of the corner reflector is divided into four panels that are controlled by servo motors so that they rotate to the same positions synchronously~\cite{goBILDA}. Similarly to the static reflectors, aluminum panels, and 80/20 aluminum extrusions were used. The reflector is shown in \autoref{fig:cornerreflectors}C. The optimal angle change for the panels was experimentally determined in \autoref{subsec:mech_mod_reflector} to maximize RCS change and modulation speed.

\subsection{Sentinel-1}
Sentinel-1 SAR data is publicly available, and one data distributor is the Alaskan Satellite Facility which was used to access data for this paper~\cite{ASF}. Sentinel-1 provides its data at different processing levels. Level 0 is the raw measured echoes of the radar represented as encoded IQ samples, while Level 1 data is already resolved into an image~\cite{SentinelDataLevels}. SARLink uses Level 0 data for processing, although Level 1 data could also theoretically be used since SAR processing preserves amplitude and phase information. 

Sentinel-1 operates at 5.405~GHz, and its primary acquisition mode is Interferometric Wide Swath (IW) mode, which uses antenna beam steering to increase its coverage area and provide more consistent resolution in the final image. IW mode is used to image most of the land on Earth, and the satellite typically revisits an area every 3-6 days. The main drawback of IW mode is that it has a reduced dwell time, only hundreds of milliseconds~\cite{Kubica_2015_TOPSAR}. IW mode achieves a range resolution of 5~m and an azimuth resolution of 20~m~\cite{sentinelResolution}. Two polarizations are measured in this mode: co-polarization (VV) and cross-polarization (VH). All data in this paper uses co-polarized measurements since corner reflectors primarily reflect incoming waves with the same linear polarization~\cite{Michelson_1995_CRPOLAR}. SARLink processes Sentinel-1 IW mode data using an existing backprojection algorithm that resolves the images and applies necessary calibrations to the data~\cite{zebker_code}. We modified this existing code base to enable subaperture processing-based backscatter communication for this work as described in \autoref{subsec:ProcessorDesign}. In addition all data is processed using the same number of pixels, despite varying image resolutions. For this paper, each pixel represents 15~m in X and 6~m in Y.
\section{SARLink Evaluation} \label{sec:evaluation}
To evaluate SARLink, we start by presenting results obtained from the theoretical analysis of the system in \autoref{sec:theoretical}. We then perform both in-lab and field experiments to evaluate each system component, including (1) field experiments to verify the performance of Sentinel-1, (2) evaluation of the mechanically modulating corner reflector in a lab, and (3) deployment of both the static and modulating reflectors in the field to evaluate the effectiveness of the SAR backscatter communication processing algorithm. 

\begin{figure}[t!]
\centering
\captionsetup{skip=1pt}
\includegraphics[width=.5\textwidth]{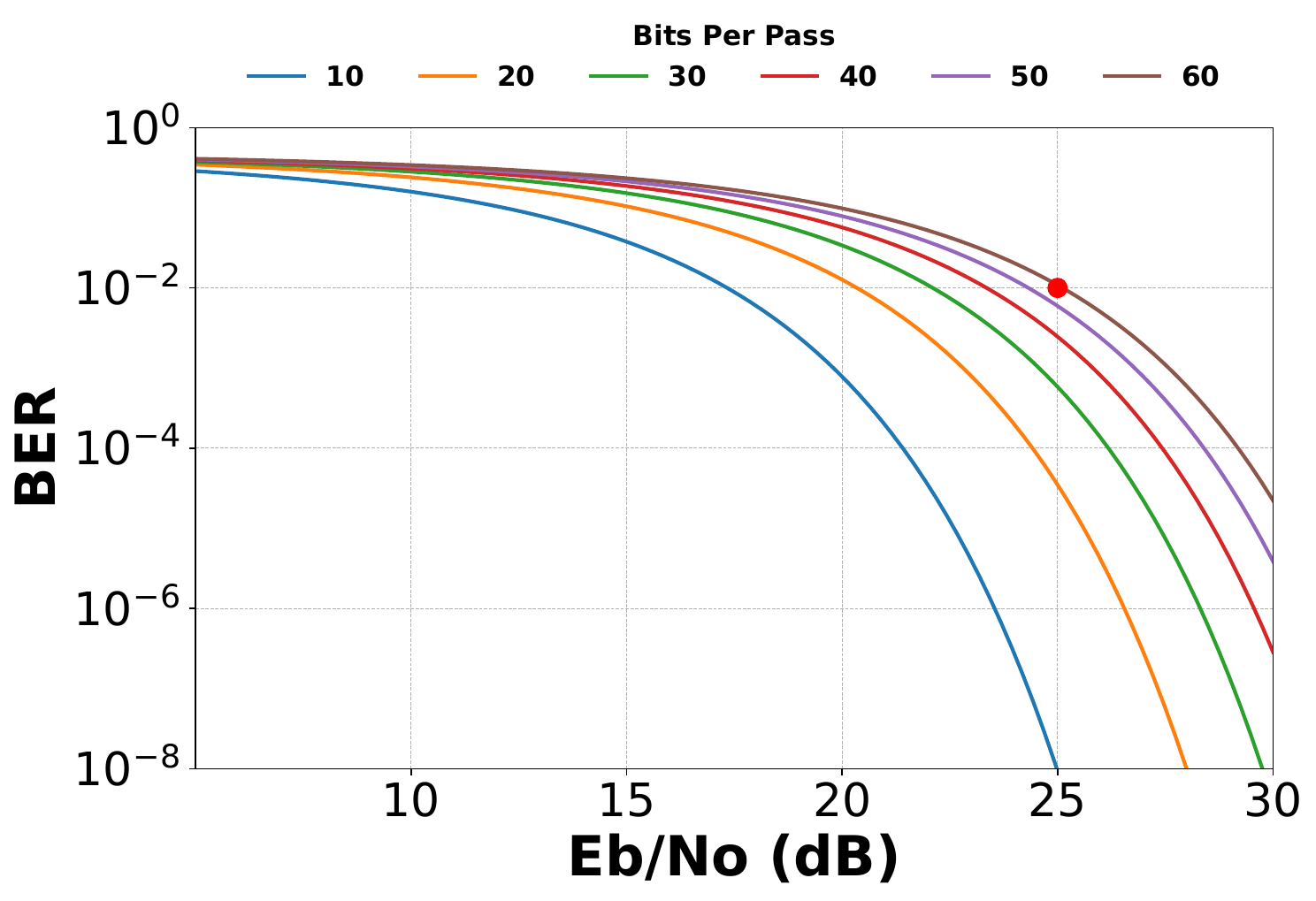}
\caption{The expected BER is plotted as a function of $\boldsymbol{E_{b}/N_{o}}$ as the total bits sent per pass is varied in each line color. The red dot marks that with $\boldsymbol{E_{b}/N_{o} = 25 dB}$,  60 bits can be sent with a 1\% BER.}
\label{fig:BER}
\Description{This plot shows the typical BER curve produced for OOK. It shows how the BER performance degrades with more bits sent.}
\end{figure}

\subsection{Theoretical Performance Evaluation} \label{subsec:theoretical_evaluation}
To evaluate the theoretical performance of SARLink, we use ~\autoref{eq:BERSAR_sigma} to determine the BER with respect to $\frac{E_{b}}{N_{o}}$. \autoref{fig:BER} shows the expected BER as $\frac{E_{b}}{N_{o}}$ increases from 5 to 30 dB, which represents half of the RCS difference between the reflector's on and off states. The plot also shows how the BER changes based on the number of sublooks used and, thus, the system's data rate. It shows the expected performance when sending 10 to 60~bits per pass.

To gain insight into the expected performance when Sentinel-1 SAR signals are explicitly used, we assume that the power difference between a 2~ft by 2~ft reflector's on and off states is 10 dB, as this was the difference we achieved using our mechanically modulating reflector (see ~\autoref{subsec:mech_mod_reflector}). Therefore, it can achieve a $\frac{E_{b}}{N_{o}}$ = 7 dB. So, while the 2~ft by 2~ft reflector is not expected to perform well with more than 10~bits per pass, using \autoref{eq:CRrcsmax}, the size of reflector required to achieve specific BER performance can be determined. For example, a 5.5~ft by 5.5~ft reflector can theoretically achieve 60~bits per pass as denoted by the red point in the plot. With this performance, the system can support the suggested applications of short messages and low-bandwidth sensor data.

\begin{figure}[t!]
\centering
\captionsetup{skip=1pt}
\includegraphics[width=0.45\textwidth]{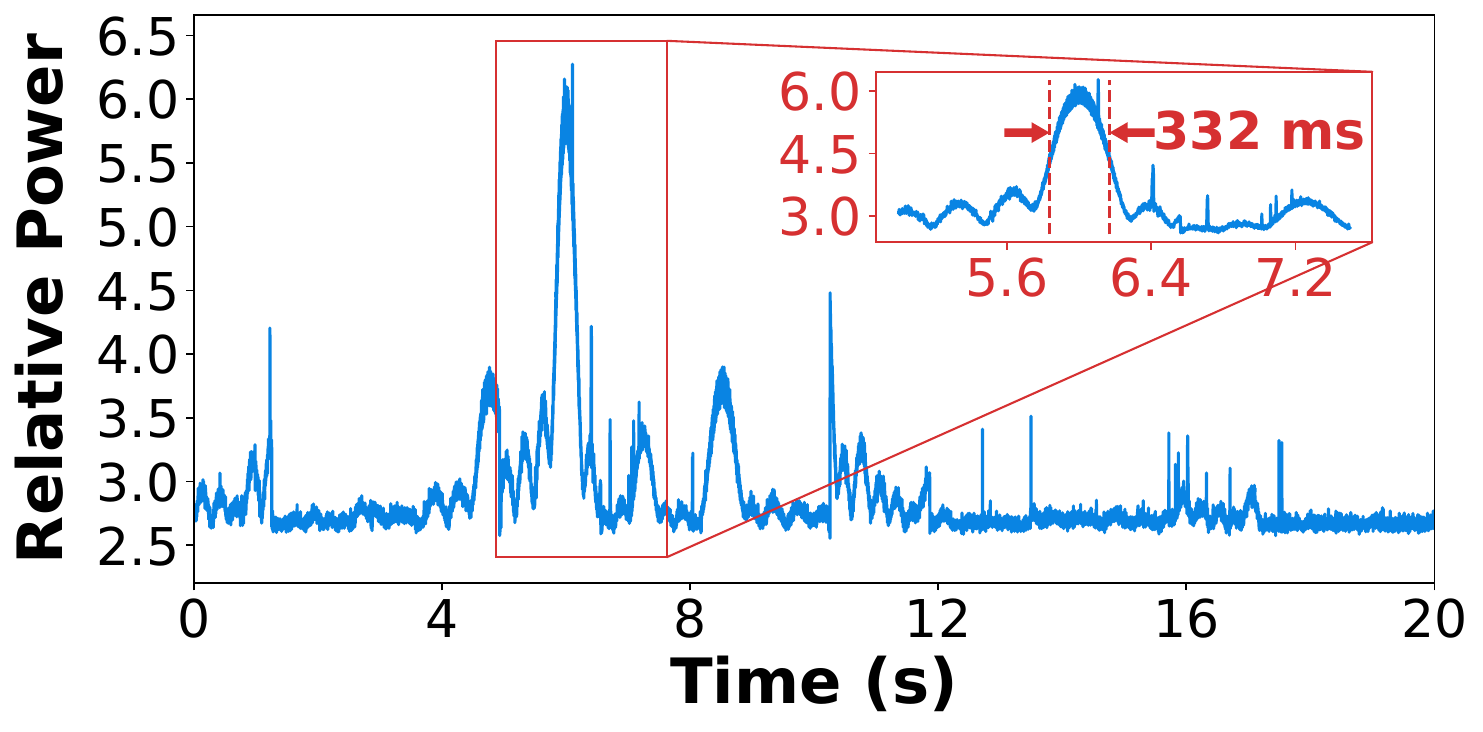}
\caption{The received power from Sentinel-1 radar signals detected on the ground using a dish antenna during imaging. The signal shows that the radar illuminated the location for 332 ms within 3 dB of the peak value.}
\label{fig:sat-pass-time-domain}
\Description{Measurements taken during Sentinel-1 imaging show the amplitude of the signal measured by the SDR at the satellite's frequency. A large peak can be seen when the area at the antenna is imaged and a zoomed in version of this plots shows that this peak lasts around 332 milliseconds.}
\end{figure}

\begin{figure*}[t!]
\centering
\captionsetup{skip=1pt}
\includegraphics[width=1\textwidth]{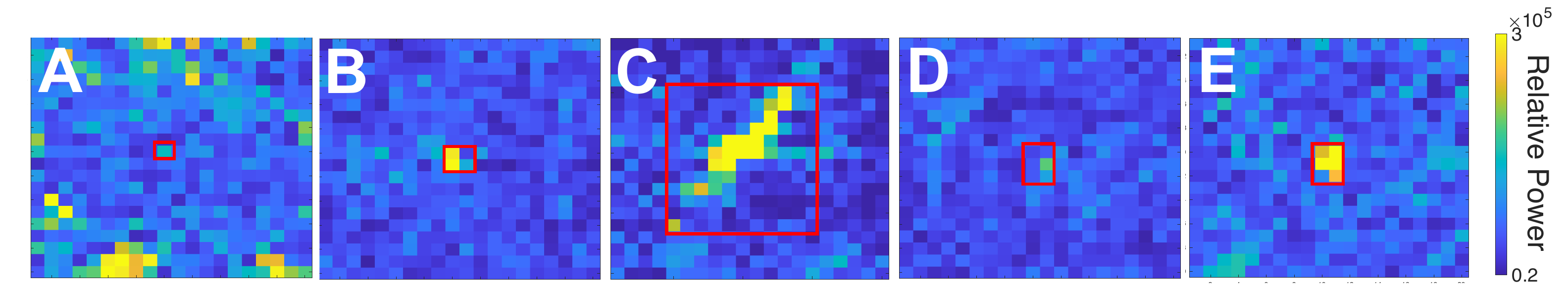}
\caption{The amplitude of corner reflectors (marked in red) pointed straight up as seen by Sentinel-1 in varying locations sized (A) 2~ft by 2~ft, (B) 3~ft by 3~ft, and (C) 4~ft by 4~ft. The  3~ft by 3~ft was also deployed facing straight up in (D) versus in (E) where it was angled towards the satellite.}
\label{fig:deployment_sizes}
\Description{Five subfigures are shown with the corner reflectors deployed as noted. The images show a blue background, the clutter, and bright yellow spots, the corner reflectors.}
\end{figure*}

\begin{figure*}[!tbp]
\centering
\begin{minipage}[b]{0.45\textwidth}
\includegraphics[width=1\textwidth]{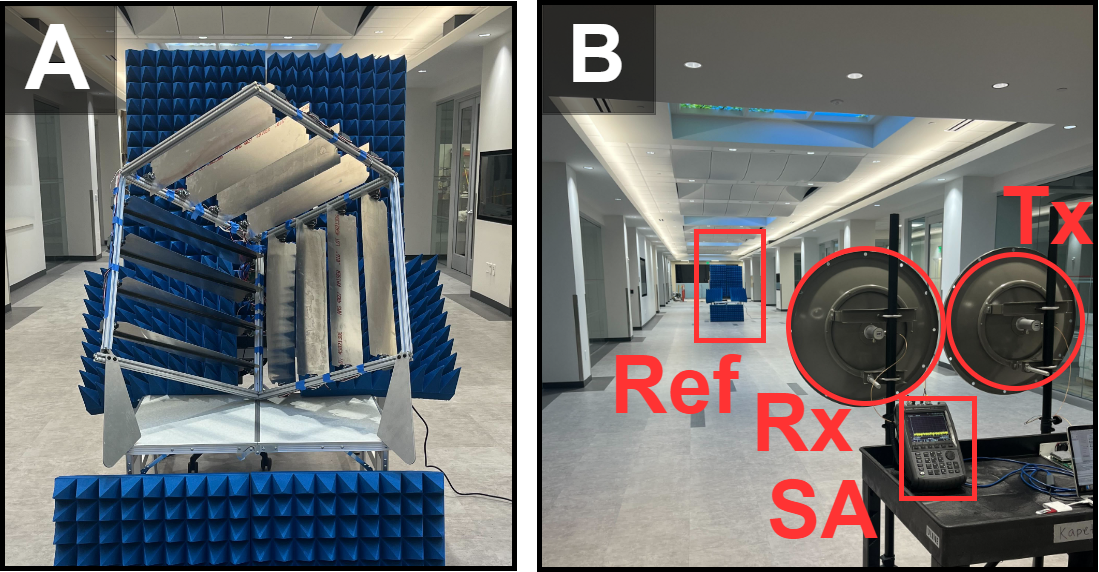}
\caption{(A) shows the corner reflector setup with absorptive foam to reduce background clutter (B) shows the test setup}
\label{fig:Mechanically_Modulating_Corner_Reflector}
\end{minipage}
\hfill
\begin{minipage}[b]{0.5\textwidth}
\centering
\includegraphics[width=1\textwidth]{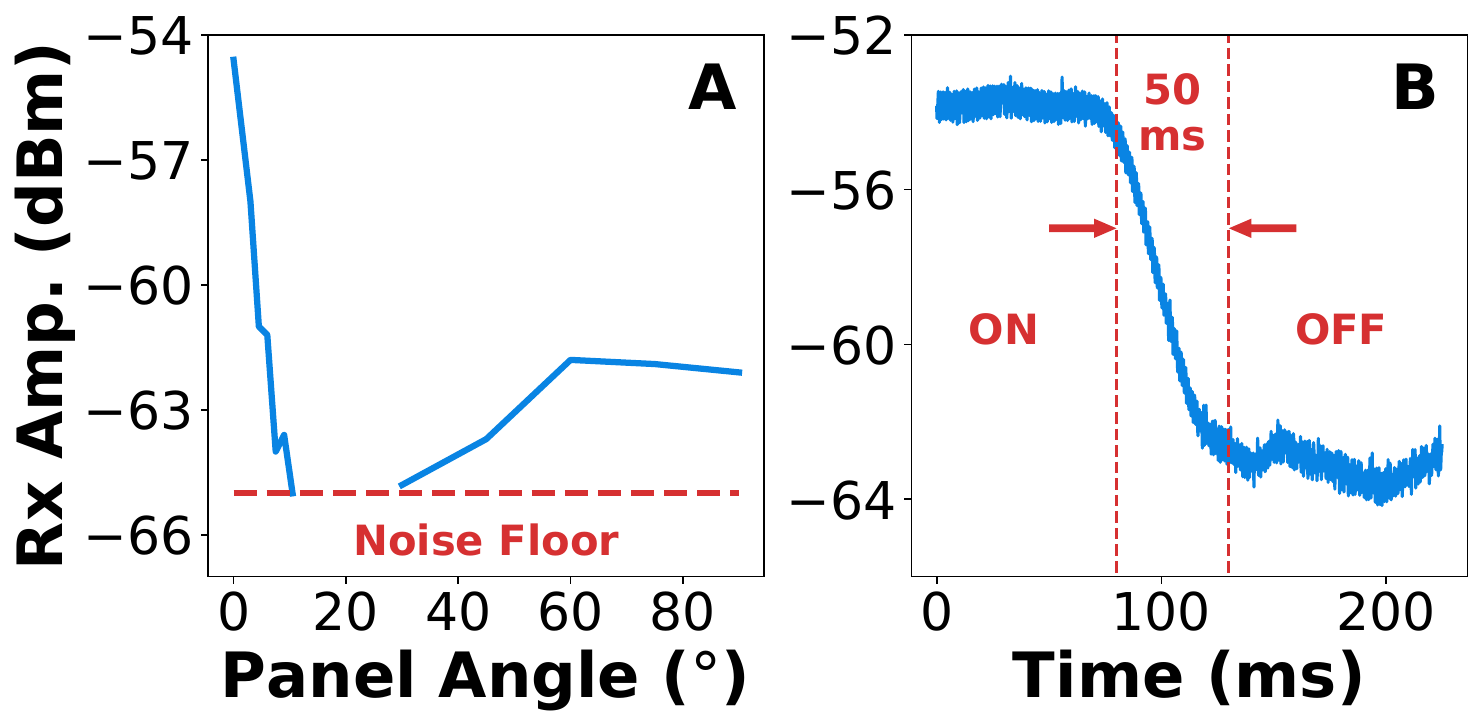}
\captionsetup{skip=1pt}
\caption{(A) The modulating mechanical reflector's reflected signal disappears below the noise floor after rotating its panels approximately 12\degree~from their flat position. (B) The maximum amplitude change between states is 10.89 dBm, and it takes ~50 ms for the reflector panels to rotate the full 12\degree.}
\label{fig:mech_ref_results}
\Description{Two two panel states of the mechanically modulating corner reflector are shown.}
\end{minipage}
\end{figure*}

\subsection{Sentinel-1 System Verification}
\label{subsec:SystemVerification}
We conducted several preliminary field experiments to ensure the Sentinel-1 SAR system functioned as expected. These experiments validated our reflector deployment method before conducting more complex evaluations.

\subsubsection{Sentinel-1 Signal Measurements} \label{subsubsec:SentinelSignal}
We recorded the satellite's transmissions to ensure the received signal arrived at the predicted time and its duration was within the expected range. \autoref{fig:sat-pass-time-domain} shows the radar signals captured with an Ettus N210 USRP software-defined radio paired with a 30~dBi dish antenna tilted and oriented towards the satellite~\cite{usrp, pasternack_datasheet}. The dwell time of the radar will depend on the range of the target. Still, it will generally be several hundred milliseconds, so our detected signal duration of 332 ms matched expectations~\cite{Kubica_2015_TOPSAR}.

\subsubsection{Static Corner Reflector Deployments} \label{subsubsec:StaticReflectorDeployment}
Next, we wanted to ensure we could reliably detect a corner reflector in the Sentinel-1 data. So, we tested several different sizes and also verified that our method of aligning the angle of the reflector would maximize its RCS as depicted in \autoref{fig:background_CR}. \autoref{fig:deployment_sizes} A, B, and C show several different-sized corner reflectors deployed in different locations. The smallest a 2~ft by 2~ft reflector takes up a single resolution cell and has low received power as compared to the 3~ft by 3~ft and 4~ft by 4~ft, which appear both bigger and brighter. We deployed these reflectors facing straight up, so their misaligned incidence angle also contributes to a lower RCS.

To resolve this issue, we deployed the same 3~ft by 3~ft corner reflector twice in the same location. \autoref{fig:deployment_sizes} D and E demonstrate the measured RCS of the reflector angled with boresight pointed straight up versus towards the satellite. The non-tilted reflector is wholly lost in the background clutter, while the tilted reflector shows up brightly. The reflector in D was imagined using a different orbital pattern than in B, which demonstrates that the required tilt angle depends on the orbit used for imaging. These experiments verified our deployment of corner reflectors for imaging by Sentinel-1, thus allowing for further experiments.

\subsection{Mechanically Modulating Reflector}\label{subsec:mech_mod_reflector}

After building the proposed mechanically modulating corner reflector, we analyzed its performance in lab to optimize the trade-off between the difference in the RCS of the reflector's on and off states and its modulation speed. This was done by running control experiments to measure the received signal strength of the backscattered signal at different panel angles and measuring the maximum achieved modulation speed using these angles before performance degradation.

\autoref{fig:Mechanically_Modulating_Corner_Reflector} shows the experimental setup used to conduct the evaluation. We placed the reflector 19~m away from a transmitter and receiver to ensure it was in the far field, and placed RF absorber foam around it to reduce clutter. The N210 USRP was connected to a 30~dBi antenna and transmitted a continuous 5.75~GHz tone (in the ISM band and near Sentinel-1's operation frequency) at -16.23~dBm~\cite{usrp}. The backscattered signal was received using a spectrum analyzer connected to an identical antenna~\cite{pasternack_datasheet}. These antennas feature a very narrow beamwidth, which minimizes crosstalk and interference despite their proximity.

In \autoref{fig:mech_ref_results}A, the measured receive power in the on-state, where the panels fully flat, is -54.6~dBm, which is slightly higher than the theoretical value of -55.8~dBm calculated from corner reflector theory~\cite{cornerreflector}. This difference is attributed to reflections from background clutter. The amplitude of received power is at its minimum (below the noise floor) when the panels are tilted at 12\degree~from the on position, creating a 10~dB contrast. The fastest the motor could switch between these states was 50~ms, plotted in \autoref{fig:mech_ref_results}B. As such, the mechanically modulating reflector is marginally able to meet the modulation speed requirements set by Sentinel-1's dwell time in IW mode.

\begin{figure*}[!tbp]
\centering
\begin{minipage}[b]{0.45\textwidth}
\includegraphics[width=1\textwidth]{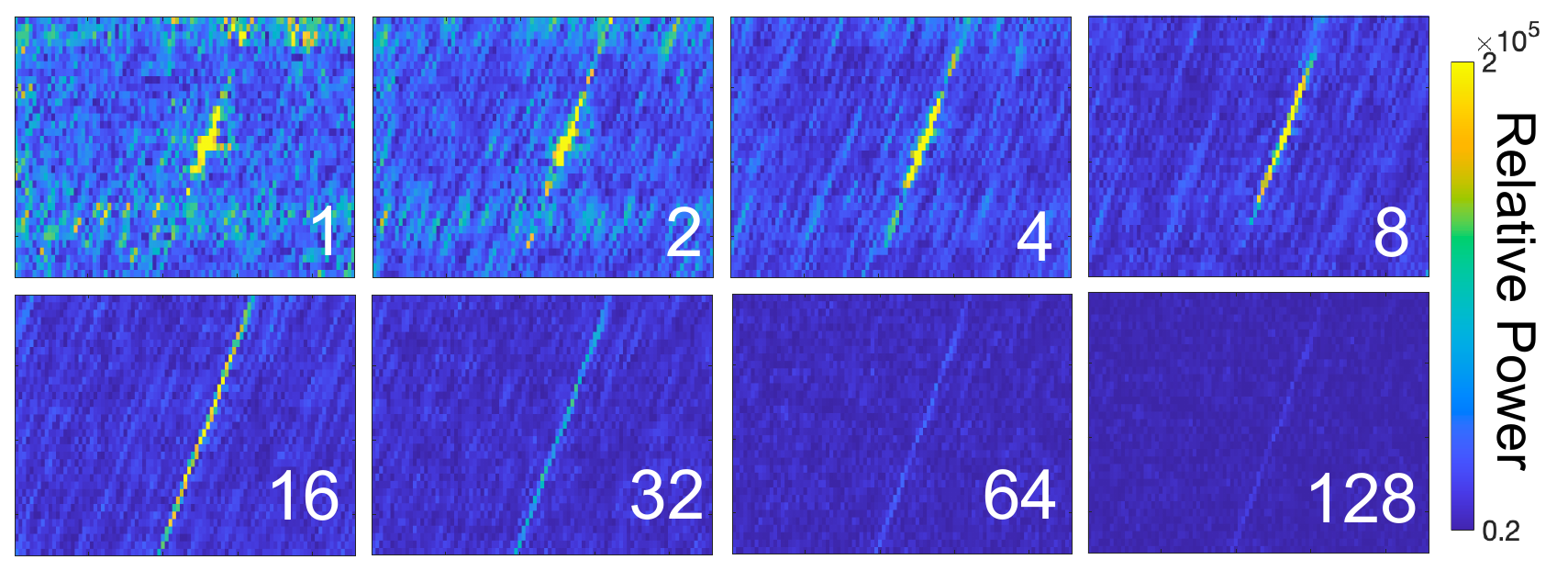}
\caption{Each image is one of the sublooks generated using m = 1, 2, 4, 8, 16, 128~subapertures as annotated on the image. The image's range and azimuth directions are not aligned to the pixel axes. The azimuth direction is tilted slightly in the clockwise direction from the vertical axis.}
\label{fig:sublooks}
\Description{The sublooks of a 4 by 4 foot corner reflector are shown. They demonstrate how the target gets stressed as the number of sublooks used increases.}
\end{minipage}
\hfill
\begin{minipage}[b]{0.5\textwidth}
\includegraphics[width=1\textwidth]{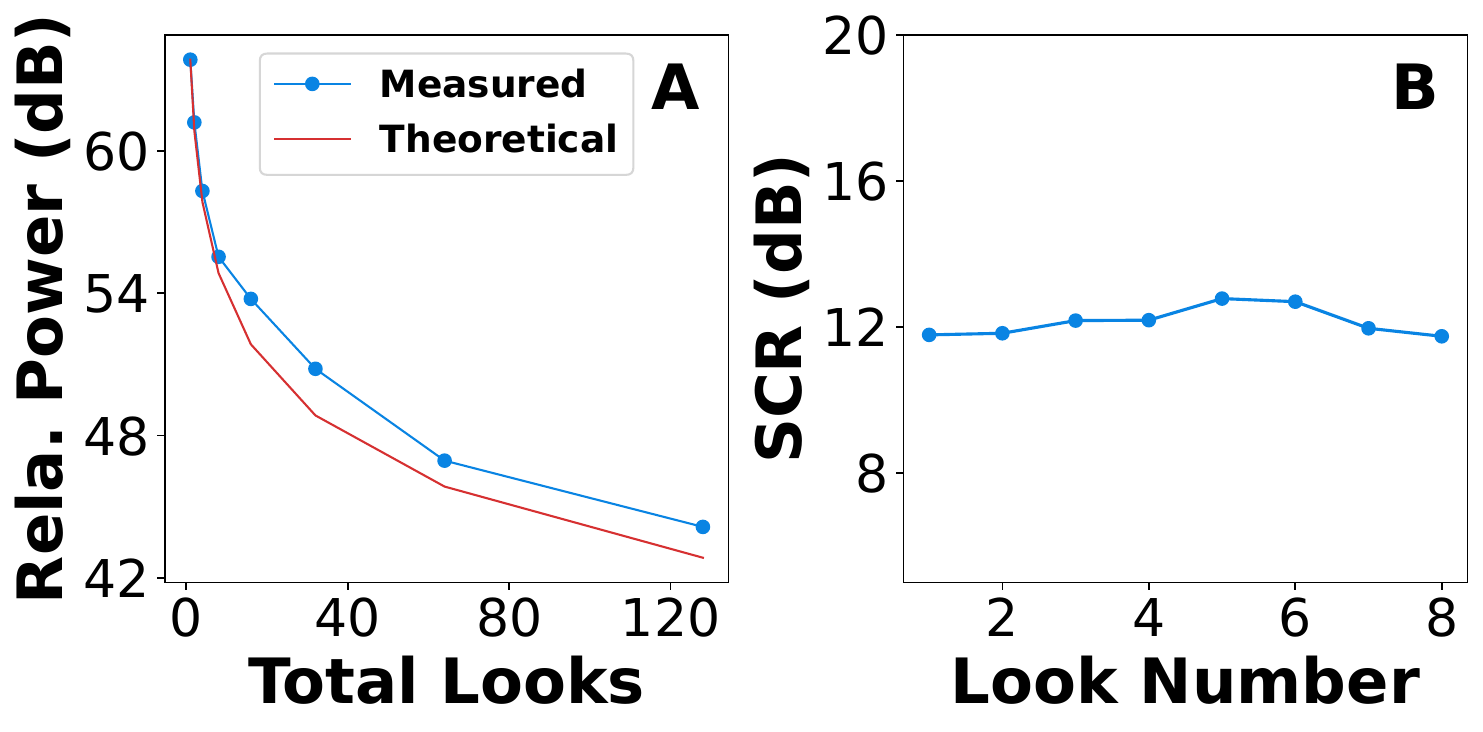}
\caption{(A) shows the relative power of the corner reflector using an increasing number of looks. The theoretical performance is shown with respect to the total number of sublooks and (B) for each of the m=8 sublooks.}
\label{fig:SCRsublooks}
\Description{There are two subfigures A and B. The plot in A shows how there is a 1/2 reduction in power every time the the number of sublooks increases by two. This is shown to be close to the theoretical expectations with a line representing the exact 1/2 reduction. The plot in B shows that each of the eight sublooks have a SCR of about 12 dB.}
\end{minipage}
\end{figure*}

\begin{figure*}[th!]
\centering
\captionsetup{skip=1pt}
\includegraphics[width=.98\textwidth]{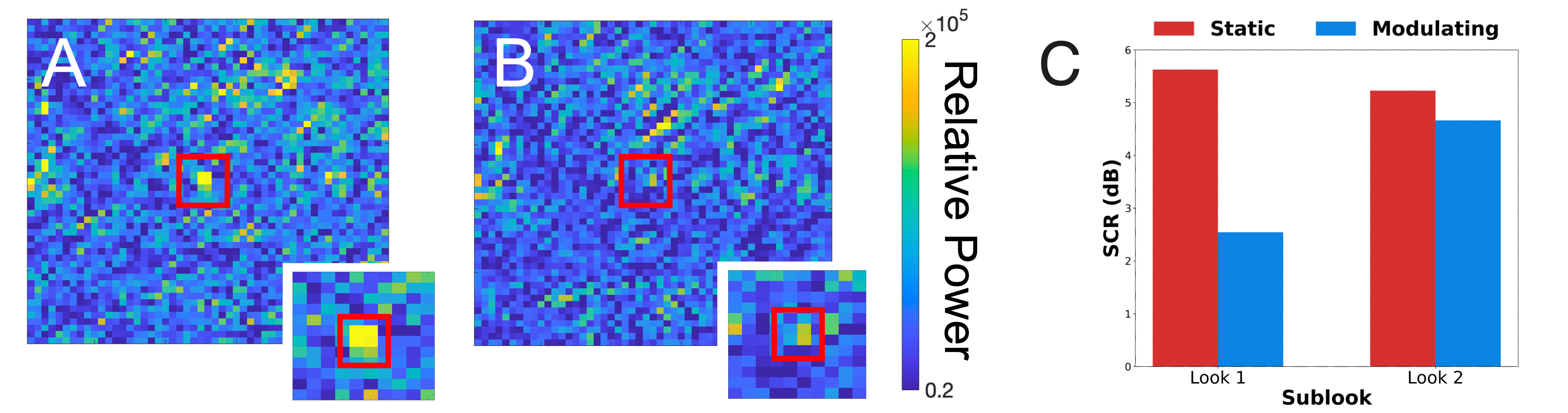}
\caption{The mechanically modulating reflector deployed \textbf{(A)} statically in its on-state and \textbf{(B)} switching with a 100 ms period. (C) SCR of Sublooks of Modulating Reflector: Static (red) and Modulating (blue). Subaperture processing of the modulating reflector shows that when its panels were turned, there was a difference in the SCR compared to when it was deployed statically.}
\label{fig:mod_ref_presence}
\Description{The first two subfigures A and B show the SAR images of the modulating reflector deployed statically and modulating. The static images is a much larger amplitude and taken up more pixels of the image. The third figure shows the results of using two sublooks to process the data and the SCR of the static is around 5 dB for both, while it is 2 dB and about 5 dB for the modulating reflector.}
\end{figure*}

\subsection{Data Processing Results} \label{sub:Processing_Results}

To evaluate the effectiveness of the subaperture data processing algorithm for backscatter communication, we assess its performance using Sentinel-1A image data with a static corner reflector and the mechanically modulating reflector.

\subsubsection{Static Reflector Processing Results} \label{subsub:static_processing_Results}
 The 4~ft by 4~ft reflector, as shown in \autoref{fig:cornerreflectors}A, was processed with m-sublooks, where m~=~1, 2, 4, 8, 16, 32, 64, and 128. In \autoref{fig:sublooks}, one of the sublooks generated for each value of m is pictured. Increasing the number of sublooks by factors of two demonstrates how doubling the azimuth resolution causes the reflector to appear increasingly stretched as more sublooks are used.

In \autoref{fig:SCRsublooks}A, we quantitatively evaluate how the total number of sublooks affects the measured reflector power. In \autoref{eq:SublookSNR}, the relative signal power reduces by a factor equal to the number of sublooks used, and this theoretical relationship is shown in red. The reflector's power is slightly higher than the predicted value because the reduced azimuth resolution averages more clutter with the reflector. Overall, these results match with expectations, validating our subaperture processing and theoretical model.

\autoref{fig:SCRsublooks}B analyzes the SCR of the eight sublook images produced by processing the reflector with eight subapertures, m=8. We evaluate in terms of SCR rather than relative power due to the beam steering used by Sentinel-1's IW mode, which causes differences in the returned power based on the sublook angle. The RCS of the reflector should have minimal variation across sublooks due to the small imaging angle used to capture the target. This is validated in the experimental results as the SCR remains constant across each of the eight sublooks. 

\subsubsection{Modulating Reflector Processing Results} \label{subsub:mod_processing_Results}
Finally, we analyzed the performance of a modulating reflector in the field. The mechanically modulating reflector was deployed twice at the same location in a large grassy field. The first time, statically in its on-state and the second time, with the panels modulating between the on and off states every 100 ms, as shown in \autoref{fig:mech_ref_results}B. The reflector location was imaged for about 100~ms for this specific deployment site, meaning there was time for a single period. The static deployment in \autoref{fig:mod_ref_presence}A has high relative receive power as compared to the background, while in \autoref{fig:mod_ref_presence}B, the same reflector modulating has a significantly reduced RCS. This test demonstrates that the reflector's changing RCS is detectable by the satellite, thus demonstrating showing that the system can detect the target's RCS state. 

We also evaluated using sublook processing to detect changes in the RCS as the satellite passed overhead. \autoref{fig:mod_ref_presence}C shows the results of using the processing code to evaluate the SCR in two sublook images.
The SCR remains relatively constant in the static deployment, around 5 dB, while the modulating reflector shows a 2 dB difference between the sublooks. This difference indicates that subaperture processing can successfully detect modulation. We deployed the reflector during the rainy season, so there was increased background clutter. Given these conditions and the size of the reflector, it was not possible to split the sublooks further without losing the reflector in the background clutter. Nevertheless, the results are a promising indication that under better deployment conditions and additional hardware optimization, the performance of SARLink can be significantly improved and evaluated further.
\section{Discussion} \label{sec:discussion}
SARLink creates a pathway to realize satellite backscatter connectivity using spaceborne SAR systems. Here, we discuss several future research directions to further improve the system established in this paper:

\textbf{Reflector Improvements:} The mechanically modulated corner reflector was built to demonstrate the feasibility of a satellite SAR-based backscatter communication system. There are several ways to improve the performance of the on-ground target, including lower power consumption and faster modulation. One way to improve power consumption is for the reflector to intelligently initialize operations to reduce to only turn on during known satellite pass times. In addition, an electronically modulating smart surface would address both reducing power consumption and increasing maximum modulation speed. Examples of this could be Van Atta arrays like those used in other backscatter works~\cite{Eid_2023_UnderwaterBackscatter}. An electronic surface could also open up opportunities to explore more sophisticated modulation schemes by allowing control of the signal's phase and polarization. Switching to an electronic design would not necessarily reduce the size of the on-ground target, but there are also approaches to make the system more practical. A promising solution would be to develop reconfigurable reflectors or smart surfaces that can maintain a small form factor until deployed~\cite{bichara2023multi}.

\textbf{Improving Throughput:} \autoref{fig:BER} demonstrated SARLink's throughput capabilities as a communication system using simple OOK. To send more data multiple reflectors can be deployed during a single pass. While the reflectors should not be placed closely in the azimuth direction, they can be very close to each other in the range direction since subaperture processing only erodes the image's azimuth resolution. Other approaches to increase performance include implementing more complex modulation schemes beyond OOK, such as quadrature amplitude modulation or by modulating both co and cross-polarizations independently to double the channel capacity. Using these approaches, the throughput of the system could likely be improved by an order of magnitude.

\textbf{Nonuniform Clutter:} In the theoretical analysis, it is assumed for simplicity that the background clutter of the scene is constant. If this system is deployed in the middle of the ocean, this assumption will hold reasonably well, but this will not be the case for deployments on land with varying terrain. In these scenarios, different sublooks will be averaged with different amounts of clutter that will change the amplitude threshold that should be used for the evaluation of the reflector state. A technique for addressing this could be using the fully resolved original SAR image to determine which clutter is averaged with each sublook and adjust the amplitude threshold accordingly.

\section{Related Work}
SARLink is related to prior work in the following domains: 

\textbf{Backscatter Communication: } SARLink is particularly related to ambient backscatter, where ambient RF signals are used to enable wireless communication, such as TV and FM broadcast signals, Wi-Fi, or BLE, and more recently, thermal noise sources~\cite{van2018ambient,kellogg2014wi, wang2017fm, ensworth2017ble, kapetanovic2023cosmic, kapetanovic2022communication}. Compared to prior work, SARLink's transmitter uses a similar design where data bits are transmitted by switching between reflective and non-reflective states. However, SARLink focuses on utilizing spaceborne SAR signals and solves several unique challenges in extracting data bits from SAR images in order to enable very long-range backscatter communication. The communication range of SARLink is orders of magnitudes greater but achieves lower bit rates compared to prior work. 

\textbf{Passive Retroreflectors:} Like SARLink, previous work has utilized systems that exploit retroreflectors for sensing and communication~\cite{Yang_2023_CubeSense, Soltanaghaei_2021_Millimetro, Zhou_2003_OpticalRetro}. The work discussed in these papers uses either mmWave or optical-based communication. They consider a much more limited range than SARLink but exploit the mechanical sensitivity of reflectors to sense and communicate in their environments.

\textbf{IoT Satellite Connectivity.} The idea of connecting remote IoT devices to satellites has been well-explored by many commercial start-ups~\cite{Swarm, starlink, HubbleNetwork}. However, active radios that transmit enough power to communicate with satellites require several Watts of power, which is orders of magnitude more than what typical energy-harvesting and battery-constrained systems could afford~\cite{paradiso2005energy}. Even state-of-the-art low-power satellite connectivity devices like Swarm M138 use upwards of 3.3 W during active transmission~\cite{Swarm}, and the cost of transmission with these devices can also be substantial (\$2.6/kB~\cite{iridium}) due to the use of private satellite networks. While optical devices can theoretically enable low-power backscatter communication in the visible light spectrum, their performance would be highly affected by cloud coverage, rendering them impractical~\cite{langhorst2024global}. This makes developing passive, low-cost communication methods with open-sourced SAR satellites such as Sentinel-1 motivating for the IoT community.

\textbf{Subaperture Processing.} The use of subaperture processing for the detection of movement in SAR images has been proposed in prior work. Sentinel-1 itself has level 2 data products that use it to investigate ocean swells \cite{ESA_2021_OceanSwell}. Prior work has also analyzed the movement of targets, including ships, by looking at the difference in the object's position between sublooks \cite{Yoshida_2021_ShipSublook, ouchi_multilook}. There has even been work done to develop SAR "video" \cite{Kim_2020_SARVideo}. However, SARLink is unique as it is the first work to propose using this processing for communication purposes.

\textbf{Modulating Reflective Surfaces.} Other works that focus on military applications, such as~\cite{WANG_SAR}, provided insights into how one might construct an active surface that selectively absorbs RF waves of specific frequencies to perform SAR jamming. They were able to achieve amplitude modulation of backscattered SAR signal using an active frequency selective surface (AFSS) constructed from bow-tie antennas and PIN diodes. However, their processing was tested in the lab with stationary radar units and simulated on measured data sets. Existing systems related to metamaterial surfaces for signal modulation are also only evaluated in simulation or in lab-scale settings~\cite{cho2022towards, arun2020rfocus, cho2021mmwall, chen2021pushing}. This makes deploying similar modulating surfaces for ground-to-satellite communication in the field a significant area of future work.

\section{Conclusion}
This paper presents SARLink, the first wireless communication system to enable passive ground-to-satellite connectivity leveraging spaceborne SAR systems. This backscatter communication system enables communication with orders of magnitude more range than traditional systems. We designed a mechanically modulating corner reflector to enable OOK with SAR signals by changing its RCS such that applied modulation is observable by a satellite in LEO. We also propose, analyze, and evaluate a data processing algorithm for SAR backscatter communication. The algorithm leverages subaperture processing to detect multiple RCS states of a target in a single satellite image, which was demonstrated in field testing with Sentinel-1A. The theoretical analysis predicts that SARLink can passively send tens of bits to SAR satellites in orbit using a single SAR image, enough to support low bandwidth sensor data and messages. This paper lays the groundwork for passive satellite connectivity for low-power sensing nodes all across Earth.

% Acknowledgements
\begin{acks}
The authors would like to thank Professor Howard Zebker for providing the Sentinel-1 image processor, Elizabeth Wig for advice regarding SAR processing, the members of the S4 lab for their helpful suggestions and feedback, and the anonymous shepherd and reviewers of this work for their comments and review of this manuscript.
\end{acks}

% Bibliography
\printbibliography

%End the Document
\end{document}